\PassOptionsToPackage{unicode}{hyperref}
\PassOptionsToPackage{hyphens}{url}
\PassOptionsToPackage{dvipsnames,svgnames,x11names}{xcolor}
\documentclass[
]{article}
\usepackage[margin=1in]{geometry}
\usepackage{amsmath,amssymb}
\usepackage{lmodern}
\usepackage{iftex}
\ifPDFTeX
  \usepackage[T1]{fontenc}
  \usepackage[utf8]{inputenc}
  \usepackage{textcomp} 
\else 
  \usepackage{unicode-math}
  \defaultfontfeatures{Scale=MatchLowercase}
  \defaultfontfeatures[\rmfamily]{Ligatures=TeX,Scale=1}
\fi
\IfFileExists{upquote.sty}{\usepackage{upquote}}{}
\IfFileExists{microtype.sty}{
  \usepackage[]{microtype}
  \UseMicrotypeSet[protrusion]{basicmath} 
}{}
\makeatletter
\@ifundefined{KOMAClassName}{
  \IfFileExists{parskip.sty}{%
    \usepackage{parskip}
  }{
    \setlength{\parindent}{0pt}
    \setlength{\parskip}{6pt plus 2pt minus 1pt}}
}{
  \KOMAoptions{parskip=half}}
\makeatother
\usepackage{xcolor}
\NewDocumentCommand\citeproctext{}{}
\NewDocumentCommand\citeproc{mm}{%
  \begingroup\def\citeproctext{#2}\cite{#1}\endgroup}
\makeatletter
 \let\@cite@ofmt\@firstofone
 \def\@biblabel#1{}
 \def\@cite#1#2{{#1\if@tempswa , #2\fi}}
\makeatother
\newlength{\cslhangindent}
\newlength{\csllabelwidth}
\newenvironment{CSLReferences}[2] 
 {\begin{list}{}{%
  \setlength{\itemindent}{0pt}
  \setlength{\leftmargin}{0pt}
  \setlength{\parsep}{0pt}
  \ifodd #1
   \setlength{\leftmargin}{\cslhangindent}
   \setlength{\itemindent}{-1\cslhangindent}
  \fi
  \setlength{\itemsep}{#2\baselineskip}}}
 {\end{list}}
\usepackage{calc}

\ifLuaTeX
\usepackage[bidi=basic]{babel}
\else
\usepackage[bidi=default]{babel}
\fi
\babelprovide[main,import]{american}

\def\languageshorthands#1{}
\ifLuaTeX
  \usepackage{selnolig}  
\fi
\IfFileExists{bookmark.sty}{\usepackage{bookmark}}{\usepackage{hyperref}}
\IfFileExists{xurl.sty}{\usepackage{xurl}}{} 
\hypersetup{
  pdftitle={adapol: Adaptive pole-fitting for quantum many-body
physics},
  pdfauthor={Zhen Huang, Chia-Nan Yeh, Lin Lin, Nils Wentzell, Jason
Kaye, Hugo U. R. Strand},
  pdflang={en-US},
  colorlinks=true,
  linkcolor={Maroon},
  filecolor={Maroon},
  citecolor={Blue},
  urlcolor={Blue},
  pdfcreator={LaTeX via pandoc}}

\title{adapol: Adaptive pole-fitting for quantum many-body physics}

\definecolor{c53baa1}{RGB}{83,186,161}
\definecolor{c202826}{RGB}{32,40,38}

\usepackage[affil-it]{authblk}
\usepackage{orcidlink}
\author[1,2,3%
  ]{Zhen Huang%
    \,\orcidlink{0000-0002-4801-8635}\,%
    }
\author[1%
  ]{Chia-Nan Yeh%
    \,\orcidlink{0000-0002-4166-0764}\,%
    }
\author[4,3,5%
  ]{Lin Lin%
    \,\orcidlink{0000-0001-6860-9566}\,%
    }
\author[1%
  ]{Nils Wentzell%
    \,\orcidlink{0000-0003-3613-007X}\,%
    }
\author[1,2]{Jason Kaye%
    \,\orcidlink{0000-0001-8045-6179}\,%
    }
\author[6%
  ]{Hugo U. R. Strand%
    \,\orcidlink{0000-0002-7263-4403}\,%
    }

\affil[1]{Center for Computational Quantum Physics, Flatiron Institute,
New York, NY 10010, USA%
  }
\affil[2]{Center for Computational Mathematics, Flatiron Institute, New
York, NY 10010, USA%
  }
\affil[3]{Department of Mathematics, University of California, Berkeley,
CA 94720, USA%
  }
\affil[4]{Department of Computing and Mathematical Sciences, California
Institute of Technology%
  }
\affil[5]{Applied Mathematics and Computational Research Division,
Lawrence Berkeley National Laboratory, Berkeley, CA 94720, USA%
  }
\affil[6]{School of Science and Technology, Örebro University, SE-70182
Örebro, Sweden%
  }
\date{}

\begin{document}
\maketitle

\section{Summary}\label{summary}

The Green's function approach to quantum many-body physics aims to
replace high-dimensional wavefunctions with correlation functions which
are more closely related to experimental observables of interest, such
as spectral and response functions. Finite temperature quantities, such
as the Green's function, self-energy, and hybridization functions, are
often represented in the discrete ``Matsubara'' domain on the imaginary
frequency axis (\citeproc{ref-Matsubara1955}{Matsubara, 1955}). A
variety of physical observables can be directly recovered from the
Matsubara Green's function, and many quantities of interest can be
calculated more efficiently in this formalism.

A common computational task within this framework is decomposing a
Matsubara function into a sum of simple poles:
\[G(\mathrm{i} \nu_n) \approx \sum_{k=1}^{M} \frac{R_k}{\mathrm{i} \nu_n-p_k}.\]
Here, \(G(\mathrm{i}\nu_{n})\) is in general an \(m \times m\)
matrix-valued function of the Matsubara frequency point
\(\mathrm{i} \nu_n = (2n+1) \pi\mathrm{i} / \beta\) for fermionic
functions, and \(\mathrm{i} \nu_n = 2 n \pi\mathrm{i} / \beta\) for
bosonic functions, with \(\beta\) representing the inverse temperature,
\(n \in \mathbb{Z}\), and \(m\) the number of quantum states or
spin-orbitals. The \(p_k\) are real-valued pole locations, and the
\(R_k\) are the corresponding matrix-valued residues. In applications
such as hybridization fitting, the poles and residues define an
effective non-interacting model, with the \(p_k\) playing the role of
energy levels, and it is often desirable to obtain an accurate fit with
as few poles as possible.

Since the pole locations enter the approximation nonlinearly and are
shared by all components of a matrix-valued function, a best fit from
Matsubara frequency data cannot be obtained component-wise, leading to a
highly non-convex optimization landscape. \texttt{adapol} (``add a
pole'') is a Python package implementing the adaptive pole-fitting
procedure outlined in (\citeproc{ref-huang2023}{Huang et al., 2023},
\citeproc{ref-huang25}{2025}). The method uses a modified version of the
AAA rational approximation algorithm
(\citeproc{ref-nakatsukasa2018}{Nakatsukasa et al., 2018}) to obtain a
guess of the pole locations \(p_k\), which can optionally be refined by
non-convex optimization. The residues \(R_k\) are then obtained by a
linear least-squares fit. This procedure has been shown to provide an
accurate and compact fit of Matsubara data in a black-box and
noise-robust manner (\citeproc{ref-huang25}{Huang et al., 2025};
\citeproc{ref-zima26}{Zima et al., 2026}).

\section{Statement of Need}\label{statement-of-need}

The ``pole-fitting'' problem described above is a crucial step in
various numerical methods, such as hybridization fitting for quantum
impurity solvers (\citeproc{ref-georges1996dynamical}{Georges et al.,
1996}) based on exact diagonalization
(\citeproc{ref-caffarel94}{Caffarel \& Krauth, 1994};
\citeproc{ref-liebsch11}{Liebsch \& Ishida, 2011};
\citeproc{ref-mejuto2020efficient}{Mejuto-Zaera et al., 2020}),
perturbation theory (\citeproc{ref-huang25}{Huang et al., 2025};
\citeproc{ref-kaye24}{Kaye et al., 2024}), and time evolution of matrix
product states (\citeproc{ref-wolf15}{Wolf et al., 2015};
\citeproc{ref-zima26}{Zima et al., 2026}), as well as certain approaches
to analytic continuation of Matsubara Green's functions
(\citeproc{ref-fei2021nevanlinna}{Fei, Yeh, \& Gull, 2021};
\citeproc{ref-fei21_2}{Fei, Yeh, Zgid, et al., 2021};
\citeproc{ref-huang2023}{Huang et al., 2023};
\citeproc{ref-ying22}{Ying, 2022a}, \citeproc{ref-ying22_2}{2022b};
\citeproc{ref-zhang24_2}{Zhang et al., 2024};
\citeproc{ref-zhang24}{Zhang \& Gull, 2024}), and other perturbation
theory-based diagrammatic methods (\citeproc{ref-gazizova24}{Gazizova et
al., 2024}, \citeproc{ref-gazizova25}{2025}). In many applications
(e.g., dynamical mean-field theory), the pole-fitting step appears
inside a self-consistent loop, requiring a black-box algorithm
delivering results with controlled accuracy.

Although significant progress has been made in the past several years on
developing algorithms to solve the pole-fitting problem
(\citeproc{ref-huang2023}{Huang et al., 2023},
\citeproc{ref-huang25}{2025};
\citeproc{ref-mejuto2020efficient}{Mejuto-Zaera et al., 2020};
\citeproc{ref-shinaoka21}{Shinaoka \& Nagai, 2021};
\citeproc{ref-ying22}{Ying, 2022a}, \citeproc{ref-ying22_2}{2022b};
\citeproc{ref-zhang24_2}{Zhang et al., 2024};
\citeproc{ref-zhang24}{Zhang \& Gull, 2024}), a lack of widely-deployed
and user-friendly software has limited the adoption of these methods.
Practitioners often still rely on ad-hoc or older, easy-to-implement
methods: for example, brute force optimization methods for pole-fitting,
or Padé approximants for analytic continuation
(\citeproc{ref-vidberg77}{Vidberg \& Serene, 1977}). \texttt{adapol}
addresses this gap by providing a simple, self-contained interface with
few user parameters, tailored for common applications.

\section{State of the field}\label{state-of-the-field}

We note two primary approaches which have recently been pursued in the
literature on the pole-fitting problem: methods based on (i) Prony's
method and its variants (\citeproc{ref-ying22}{Ying, 2022a},
\citeproc{ref-ying22_2}{2022b}; \citeproc{ref-zhang24_2}{Zhang et al.,
2024}; \citeproc{ref-zhang24}{Zhang \& Gull, 2024}), and (ii) AAA
rational approximation (\citeproc{ref-nakatsukasa2018}{Nakatsukasa et
al., 2018}) followed by non-convex optimization
(\citeproc{ref-huang2023}{Huang et al., 2023},
\citeproc{ref-huang25}{2025}). The MiniPole Python package
(\citeproc{ref-minipole}{Zhang et al., 2025}) implements the Prony's
method-based approach described in (\citeproc{ref-zhang24_2}{Zhang et
al., 2024}; \citeproc{ref-zhang24}{Zhang \& Gull, 2024}), while
\texttt{adapol} implements the AAA-based approach described in
(\citeproc{ref-huang2023}{Huang et al., 2023},
\citeproc{ref-huang25}{2025}). These methods are distinct, and the
availability of both packages will allow users to compare the two
approaches.

\section{Software design}\label{software-design}

\texttt{adapol} is a simple and self-contained package which can be
incorporated into codes requiring Matsubara pole-fitting. Users can
provide Matsubara data, or an existing pole expansion to be compressed;
for example, a discrete Lehmann representation (DLR)
(\citeproc{ref-kaye2022discrete}{Kaye et al., 2022}). They can choose to
perform the fit with or without optimization-based post-processing of
the AAA result, and can specify either a maximum number of poles or a
target error tolerance. \texttt{adapol} functions are documented
extensively both within the API reference documentation, and in example
notebooks demonstrating various use cases and modes of operation. An
interface to the TRIQS package
(\citeproc{ref-parcollet2015triqs}{Parcollet et al., 2015}) is also
provided.

\section{Research impact statement}\label{research-impact-statement}

In the context of analytic continuation, the AAA-based approach
described in (\citeproc{ref-huang2023}{Huang et al., 2023}) is
considered as one of the state-of-the-art methods, and has been cited
extensively. For many other pole-fitting applications, reducing the
number of poles required to achieve a given accuracy as much as possible
is often crucial, as computational costs often scale exponentially with
the number of poles. This is the case, for example, in exact
diagonalization quantum impurity solvers
(\citeproc{ref-caffarel94}{Caffarel \& Krauth, 1994};
\citeproc{ref-liebsch11}{Liebsch \& Ishida, 2011};
\citeproc{ref-mejuto2020efficient}{Mejuto-Zaera et al., 2020}) and
diagrammatic evaluation methods (\citeproc{ref-gazizova24}{Gazizova et
al., 2024}, \citeproc{ref-gazizova25}{2025};
\citeproc{ref-huang25}{Huang et al., 2025}; \citeproc{ref-kaye24}{Kaye
et al., 2024}). Recent developments in tensor network-based quantum
impurity solvers (\citeproc{ref-zima26}{Zima et al., 2026}) also benefit
substantially from compact pole approximations of the hybridization
function, with a significant increase in computational cost observed as
the size of this approximation grows. The \texttt{adapol} algorithm was
shown in (\citeproc{ref-huang25}{Huang et al., 2025}) to consistently
yield a more compact pole approximation of a fixed, given Green's
function than the generic DLR approach
(\citeproc{ref-kaye2022discrete}{Kaye et al., 2022}), which is itself an
exponential-in-\(\beta\) improvement over naive uniform frequency grid
approaches.

\section{AI usage disclosure}\label{ai-usage-disclosure}

Generative AI tools such as Claude and Codex were used to assist in
writing code, tests, documentation, and examples in the \texttt{adapol}
package. The content produced by these tools was reviewed and edited by
the authors.

\section{Acknowledgements}\label{acknowledgements}

This work is partially supported by the Simons Targeted Grants in
Mathematics and Physical Sciences on Moiré Materials Magic (Z.H., L.L.).
H.U.R.S acknowledges financial support from the Swedish Research Council
(Vetenskapsrådet, VR) grant number 2024-04652 and funding from the
European Research Council (ERC) under the European Union's Horizon 2020
research and innovation programme (grant agreement No.~854843-FASTCORR).
The Flatiron Institute is a division of the Simons Foundation.

\section*{References}\label{references}
\addcontentsline{toc}{section}{References}

\protect\phantomsection\label{refs}

\end{document}